\documentclass[aps,twocolumn,amsmath,letterpaper]{revtex4}
\usepackage{amssymb}
\usepackage{graphicx}
\usepackage{array}
\usepackage{hhline}
\usepackage{longtable}

\newcommand\TopSpaceInTable{\rule{0pt}{2.6ex}}
\newcommand\BotSpaceInTable{\rule[-1.2ex]{0pt}{0pt}}

\begin{document}

    \title{Phase Separation and Charge-Ordered Phases of the $d = 3$ Falicov-Kimball Model\\
    at T$>$0: Temperature-Density-Chemical Potential Global Phase Diagram from Renormalization-Group Theory}

    \author{Ozan S. Sar\i yer$^{1}$, Michael Hinczewski$^{2,3}$, and A. Nihat Berker$^{4,5}$}
    \affiliation{$^1$Department of Physics, Ko\c{c} University, Sar\i yer 34450, Istanbul, Turkey,}
    \affiliation{$^2$Feza G\"ursey Research Institute, T\"UB\.ITAK - Bosphorus University, \c{C}engelk\"oy 34680, Istanbul, Turkey,}
    \affiliation{$^3$Institute for Physical Science and Technology, University of Maryland, College Park, Maryland 20742, U.S.A.,}
    \affiliation{$^4$Faculty of Engineering and Natural Sciences, Sabanc\i~University, Orhanl\i , Tuzla 34956, Istanbul, Turkey,}
    \affiliation{$^5$Department of Physics, Massachusetts Institute of Technology, Cambridge, Massachusetts 02139, U.S.A.}

\begin{abstract}

The global phase diagram of the spinless Falicov-Kimball model in $d
= 3$ spatial dimensions is obtained by renormalization-group theory.
This global phase diagram exhibits five distinct phases. Four of
these phases are charge-ordered (CO) phases, in which the system
forms two sublattices with different electron densities.  The CO
phases occur at and near half filling of the conduction electrons
for the entire range of localized electron densities.  The phase
boundaries are second order, except for the intermediate and large
interaction regimes, where a first-order phase boundary occurs in
the central region of the phase diagram, resulting in phase
coexistence at and near half filling of both localized and
conduction electrons.  These two-phase or three-phase coexistence
regions are between different charge-ordered phases, between
charge-ordered and disordered phases, and between dense and dilute
disordered phases. The second-order phase boundaries terminate on
the first-order phase transitions via critical endpoints and double
critical endpoints. The first-order phase boundary is delimited by
critical points. The cross-sections of the global phase diagram with
respect to the chemical potentials and densities of the localized
and conduction electrons, at all representative interactions
strengths, hopping strengths, and temperatures, are calculated and
exhibit ten distinct topologies.

PACS numbers: 71.10.Hf, 05.30.Fk, 64.60.De, 71.10.Fd

\end{abstract}

    \maketitle
    \def\s{\rule{0in}{0.28in}}
    \setlength{\LTcapwidth}{\columnwidth}

\section{Introduction}

\begin{figure*}[]
\centering
\includegraphics*[scale=1]{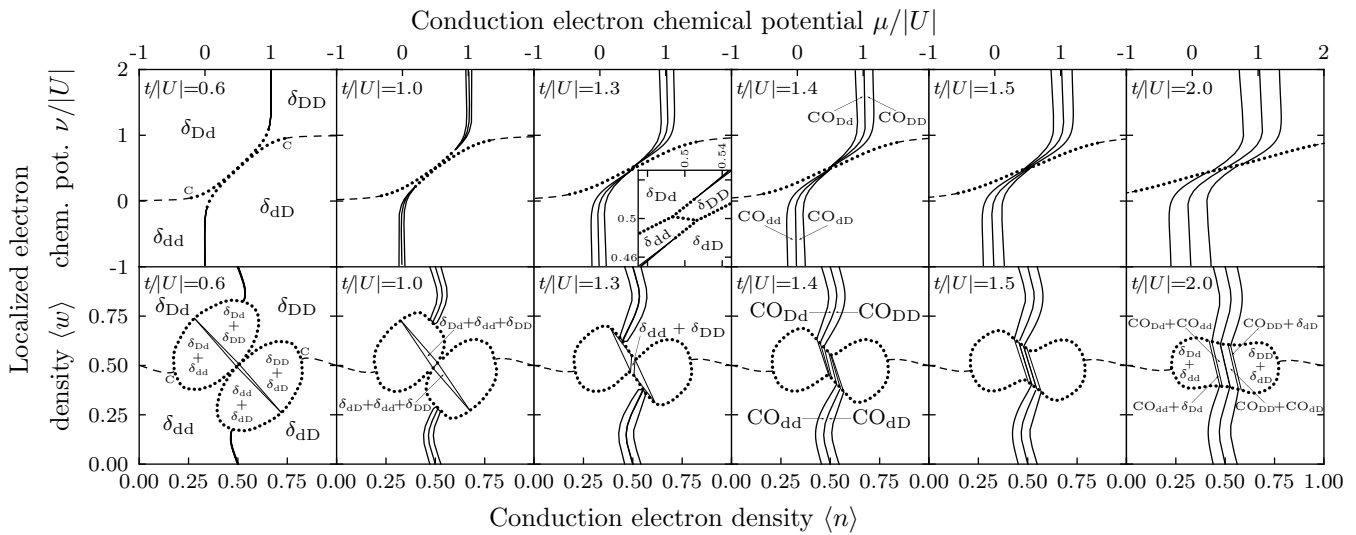}
\caption{Evolution of the cross-sections of the global phase diagram
under $t$ increase for $|U|=1$, in terms of the chemical potentials
(upper panels) and densities (lower panels) of the localized and
conduction electrons. In phase subscripts throughout this paper, the
first and second subscripts respectively describe localized and
conduction electron densities, as dilute (d) or dense (D).  The
dotted and thick full lines are respectively first- and second-order
phase transitions.  Phase separation, \textit{i.e.}, phase
coexistence occurs inside the dotted boundaries, as identified
appropriately but not repeatedly in the figure.  The dashed lines
are not phase transitions, but smooth changes between the different
density regions of the disordered ($\delta$) phase.  The
charge-ordered phases are denoted by CO.  The charge-ordered phases
occur as strips near the half filling of the conduction electrons.
Phase separation occurs near the simultaneous half filling of both
the localized and conduction electrons. The four rounded coexistence
regions, two of which disappear as $t$ is increased, are two-phase
coexistence regions between the disordered $\delta$ phases that are
distinguished by electron densities.  The narrow triangular regions
are three-phase coexistence regions between the $\delta$ phases. The
second-order transition lines bounding the charge-ordered CO phases
terminate at critical endpoints on the coexistence regions.  These
endpoints, as $t$ is increased, move past each other, as detailed in
Sec.IV below, leading to coexistence regions between the different
charge-ordered phases and between the charge-ordered and disordered
phases.}
\end{figure*}

The Falicov-Kimball model (FKM) was first proposed by L. M. Falicov
and Kimball \cite{Falicov} to analyze the thermodynamics of
semiconductor-metal transitions in SmB$_6$ and transition-metal
oxides \cite{RamirezFalicov70, RamirezFalicov71, Plischke,
GoncalvesDaSilva}. The model incorporates two types of electrons:
one type can undergo hopping between sites and the other type cannot
hop, thereby being localized at the sites. Thus, in its
introduction, FKM described the Coulomb interaction between mobile
$d$ band electrons and localized $f$ band electrons. There have been
a multitude of subsequent physical interpretations based on this
interaction, including that of localized ions attractively
interacting with mobile electrons, which yields crystalline
formation \cite{Kennedy, Lieb}. Another physical interpretation of
the model is as a binary alloy, in which the localized degree of
freedom reflects A or B atom occupation \cite{FreericksFalicov,
Freericks}. In this paper we employ the original language, with $d$
and $f$ electrons as conduction and localized electrons with a
repulsive interaction between them.

Since there is no interacting spin degree of freedom in the
Hamiltonian, the model is traditionally studied in the spinless
case, commonly referred as the spinless FKM (SFKM) and which is in
fact a special case of the Hubbard model in which one type of spin
(\emph{e.g.}, spin-up) cannot hop \cite{Hubbard}. In spite of its
simplicity, this model is able to describe many physical phenomena
in rare-earth and transition metal compounds, such as metal
transitions, charge ordering, \emph{etc}.

Beyond the introduction of the spin degree of freedom for both
electrons \cite{Brandt, Farkasovsky99, Zlatic, Minh-Tien, Lemanski,
Farkasovsky05, BrydonGulacsi06a, ZlaticFreericks01a,
ZlaticFreericks01b, ZlaticFreericks03, MillerFreericks01,
FreericksNikolic01, FreericksNikolic02, AllubAlascio96,
AllubAlascio97, LetfulovFreericks01, Ramakrishnan03}, there also
exist many extensions of the original model. The most widely studied
extensions include multiband hybridization \cite{Lawrence, Leder,
Hanke, Baeck, LiuHo83, LiuHo84, Portengen96, BrydonGulacsi06b},
$f-f$ hopping \cite{Kotecky, Batista, Yin03, Batista04, Ueltschi},
correlated hopping \cite{Michielsen, Tranh-Hai,Gajek, Wojtkiewicz01,
CencarikovaFarkasovsky04}, non-bipartite lattices
\cite{CencarikovaFarkasovsky07, GruberMacris97}, hard-core bosonic
particles \cite{GruberMacris97}, magnetic fields
\cite{ZlaticFreericks01b, AllubAlascio96, AllubAlascio97,
LetfulovFreericks01, Ramakrishnan03, GruberMacris97,
FreericksZlatic98}, and next-nearest-neighbor hopping
\cite{Wojtkiewicz06}. Exhaustive reviews are available in Refs.
\cite{FreericksZlatic03, Jedrzejewski, GruberMacris96,
GruberUeltschi}.  The wider physical application of both the basic
FKM and its extended versions have aimed at explaining valence
transitions \cite{Farkasovsky99, ZlaticFreericks01a,
ZlaticFreericks01b, ZlaticFreericks03}, metal-insulator transitions
\cite{Farkasovsky99, MillerFreericks01, FreericksNikolic01,
FreericksNikolic02, FreericksNikolic03}, mixed valence phenomena
\cite{SubrahmanyamBarma88}, Raman scattering
\cite{FreericksDevereaux01}, colossal magnetoresistance
\cite{AllubAlascio96, AllubAlascio97, LetfulovFreericks01,
Ramakrishnan03}, electronic ferroelectricity \cite{Batista,
Batista04, Portengen96, Yin03}, and phase separation
\cite{Farkasovsky99, Ueltschi, Michielsen, Freericks96,
Freericks02a, Freericks02b}.

After the initial works on the FKM \cite{Falicov, RamirezFalicov70,
RamirezFalicov71, Plischke, GoncalvesDaSilva}, the literature had to
wait 14 years for the celebrated first rigorous results. Two
independent studies, by Kennedy and Lieb \cite{Kennedy, Lieb} and by
Brandt and Schmidt \cite{BrandtSchmidt86, BrandtSchmidt87}, proved
for dimensions $d\geq 2$ that, at low temperatures, FKM has
long-range charge order with the formation of two sublattices.
Various methods have been used in the study of the FKM. In most of
these studies, either the $d\rightarrow\infty$ infinite-dimensional
limit or $d=1, 2$ low-dimensional cases have been investigated.
Studies include limiting cases such as ground-state analysis or the
large interaction limit. Renormalization-group theory \cite{Wilson}
offers fully physical and fairly easy techniques to yield global
phase diagrams and other physical phenomena.

This non-trivial nature of SFKM motivated us to determine the global
phase diagram of the model, which resulted in a richly complex phase
diagram involving charge ordering and phase coexistence, as
exemplified in Fig. 1.

We use the general method for arbitrary dimensional quantum systems
developed by A. Falicov and Berker \cite{AFalicov} to obtain the
global phase diagram of the SFKM in $d=3$, in terms of both the
chemical potentials and the densities of the two types of electrons,
for all temperatures.  The outline of this paper is as follows: In
Sec.II, we introduce the SFKM and, in Sec.III, we present the method
\cite{AFalicov}. Calculated phase diagrams are presented in Sec.IV,
for the non-hopping ($t=0$) classical submodel and for the hopping
($t\neq0$) quantum regimes of small, intermediate, and large $|U|$.
We conclude the paper in Sec.V.

\section{Spinless Falicov-Kimball Model}

The SFKM is defined by the Hamiltonian
    \begin{equation}
        \label{eq:1}
        \begin{split}
            -\beta \mathcal{H}&=t\sum_{\langle ij \rangle} \left(c_i^{\dag}c_j+c_j^{\dag}c_i\right)+U_0\sum_{i}n_iw_i\\
            &+\mu_0\sum_{i}n_i+\nu_0\sum_{i}w_i\:,
        \end{split}
    \end{equation}
where $\beta=1/k_{B}T$ and $\langle ij \rangle$ denotes that the sum
runs over all nearest-neighbor pairs of sites.  Note that, as in all
renormalization-group studies, the Hamiltonian has absorbed the
inverse temperature.  The dimensionless hopping strength $t$ can
therefore be used as the inverse temperature. Here $c_{i}^{\dag}$
and $c_i$ are respectively creation and annihilation operators for
the conduction electrons at lattice site $i$, obeying the
anticommutation rules $\{c_i,c_j\}=\{c_i^{\dag},c_j^{\dag}\}=0$ and
$\{c_i^{\dag},c_j\}=\delta_{ij}$, while $n_i=c_i^{\dag}c_i$ and
$w_i$ are electron number operators for conduction and localized
electrons respectively. The operator $w_i$ takes the values $1$ or
$0$, for site $i$ being respectively occupied or unoccupied by a
localized electron.  The particles are fermions, so that the Pauli
exclusion principle forbids the occupation of a given site by more
than one localized electron or by more than one conduction electron.

The first term of the Hamiltonian is the kinetic energy term,
responsible for the quantum nature of the model. The system being
invariant under sign change of $t$ (via a phase change of the local
basis states in one sublattice), only positive $t$ values are
considered.  The second term is the screened on-site Coulomb
interaction between localized and conduction electrons, with
positive and negative $U_0$ values corresponding to attractive and
repulsive interactions.  We consider only the repulsive case, since
the attractive case can be connected to the repulsive one by the
particle-hole symmetry possessed by either type of electrons.
Particle-hole symmetries are achieved by the transformations of
$w_i\rightarrow1-w_i$ for the localized electrons and
$c_i^{\dag}\rightarrow\kappa_ic_i$,
$c_i\rightarrow\kappa_ic_i^{\dag}$ for the conduction electrons,
where, for a bipartite lattice, $\kappa_i=1$ for one sublattice and
$\kappa_i=-1$ for the other~\cite{FreericksFalicov,GruberUeltschi}.
The last two terms of the Hamiltonian are the chemical potential
terms with $\nu_0$ and $\mu_0$ being the chemical potential for a
localized and conduction electron.

In order to carry out a renormalization-group transformation easily,
we trivially rearrange the Hamiltonian given in Eq.(\ref{eq:1}) into
the equivalent form of
    \begin{equation}
        \label{eq:2}
        \begin{split}
            -\beta\mathcal{H}=&\sum_{\langle ij\rangle} \bigg[t\left(c_i^{\dag}c_j+c_j^{\dag}c_i\right)+U\left(n_iw_i+n_jw_j\right)\\
            &\qquad\quad\quad~~\,+\mu\left(n_i+n_j\right)+\nu\left(w_i+w_j\right)\bigg]\\
            \equiv&\sum_{\langle ij\rangle}\left[-\beta\mathcal{H}_{i,j}\right],
        \end{split}
    \end{equation}
where, for a $d$-dimensional hypercubic lattice, $U=U_0/2d$,
$\mu=\mu_0/2d$, $\nu=\nu_0/2d$, and $-\beta\mathcal{H}_{i,j}$ is the
two-site Hamiltonian involving only nearest-neighbor sites $i$ and $j$.

    \section{Renormalization-Group Theory}

    \subsection{Suzuki-Takano Method in $d=1$}

    In $d=1$ the Hamiltonian in Eq.(\ref{eq:2}) is
    \begin{equation}
        \label{eq:3}
        -\beta\mathcal{H}=\sum_i\left[-\beta\mathcal{H}_{i,i+1}\right].
    \end{equation}
\noindent The renormalization-group procedure traces out half of the
degrees of freedom in the partition function \cite{Suzuki,Takano},
    \begin{equation}
        \label{eq:4}
        \begin{split}
            \mbox{Tr}_{\mbox{\tiny odd}}e^{-\beta\mathcal{H}}=&\mbox{Tr}_{\mbox{\tiny odd}}e^{\sum_i\left[-\beta \mathcal{H}_{i,i+1}\right]}\\
            =&\mbox{Tr}_{\mbox{\tiny odd}} e^{\sum_i^{\mbox{\tiny odd}}\left[-\beta\mathcal{H}_{i-1,i}-\beta\mathcal{H}_{i,i+1}\right]}\\
            \simeq&\prod_i^{\mbox{\tiny odd}}\mbox{Tr}_ie^{\left[-\beta\mathcal{H}_{i-1,i}-\beta\mathcal{H}_{i,i+1}\right]}\\
            =&\prod_{i}^{\mbox{\tiny odd}}e^{-\beta^{\prime }\mathcal{H}^{\prime}_{i-1,i+1}}\\
            \simeq&e^{\sum_i^{\mbox{\tiny odd}}\left[-\beta^{\prime}\mathcal{H}^{\prime}_{i-1,i+1}\right]}=e^{-\beta^{\prime}\mathcal{H}^{\prime}}.
        \end{split}
    \end{equation}
Here and throughout this paper primes are used for the renormalized
system. Thus, as an approximation, the non-commutativity of the
operators beyond three consecutive sites is ignored at each
successive length scale, in the two steps indicated by $\simeq$ in
the above equation.  Earlier studies \cite{AFalicov, Suzuki, Takano,
Tomczak, TomczakRichter96, TomczakRichter03, Hinczewski05,
Hinczewski06, Hinczewski08, Kaplan2, Kaplan, Sariyer} have
established the quantitative validity of this procedure.

    The above transformation is algebraically summarized in
    \begin{equation}
        \label{eq:5}
        e^{-\beta^{\prime}\mathcal{H}^{\prime}_{i,k}}=\mbox{Tr}_{\!\tiny j\,}e^{\left\{-\beta\mathcal{H}_{i,j}-\beta\mathcal{H}_{j,k}\right\}}\:,
    \end{equation}
where $i,j,k$ are three successive sites. The operator
$-\beta^{\prime}\mathcal{H}^{\prime}_{i,k}$ acts on two-site states,
while the operator $-\beta\mathcal{H}_{i,j}-\beta \mathcal{H}_{j,k}$
acts on three-site states. Thus we can rewrite Eq.(\ref{eq:5}) in
matrix form as
    \begin{multline}
        \label{eq:6}
        \langle u_iv_k|e^{-\beta^{\prime}\mathcal{H}^{\prime}_{i,k}}|\bar{u}_i^{{}}\bar{v}_k^{{}}\rangle=\\
        \sum_{s_j}\langle u_i\,s_j\,v_k|e^{-\beta\mathcal{H}_{i,j}-\beta\mathcal{H}_{j,k}}|\bar{u}_i\,s_j\,\bar{v}_k^{{}}\rangle,
    \end{multline}
where state variables $u_\ell$, $v_\ell$, $s_\ell$, $\bar{u}_\ell$,
and $\bar{v}_\ell$ can be one of the four possible single-site
$|w_\ell,n_\ell\rangle$ states at each site $\ell$, namely one of
$|00\rangle$, $|01\rangle$, $|10\rangle$, and $|11\rangle$.
Eq.(\ref{eq:6}) indicates that the unrenormalized $64\times64$
matrix on the right-hand side is contracted into the renormalized
$16\times16$ matrix on the left-hand side. We use two-site basis
states, $\{|\phi_{p}\rangle\}$, and three-site basis states,
$\{|\psi_{q}\rangle\}$, in order to block-diagonalize the matrices
in Eq.(\ref{eq:6}). These basis states are the eigenstates of total
localized and conduction electron numbers. The set of
$\{|\phi_{p}\rangle\}$ and $\{|\psi_{q}\rangle\}$ are given in
Tables \ref{tab:1} and \ref{tab:2} respectively. The corresponding
block-diagonal Hamiltonian matrices are given in Appendices A and B.

\begin{table}[h]
    \begin{tabular}{c c c c}
        \hline
        \hline
        $w$~~~~ & ~~~~~$n$~~~~~ & ~~~~~$u$~~~~~ & Two-site basis states\\
        \hline
        $0$~~~~ & $0$ & $+$ &~~~~~~$|\phi_{1}\rangle=|00,00\rangle$\\
        $0$~~~~ & $1$ & $+$ &~~~~~~$|\phi_{2}\rangle=\frac{1}{\sqrt{2}}\{|00,01\rangle+|01,00\rangle\}$\\
        $0$~~~~ & $1$ & $-$ &~~~~~~$|\phi_{3}\rangle=\frac{1}{\sqrt{2}}\{|00,01\rangle-|01,00\rangle\}$\\
        $0$~~~~ & $2$ & $-$ &~~~~~~$|\phi_{4}\rangle=|01,01\rangle$\\
        \hline
        $1$~~~~ & $0$ & $+$ &~~~~~~$|\phi_{5}\rangle=|00,10\rangle$\\
        $1$~~~~ & $1$ & $+$ &~~~~~~$|\phi_{6}\rangle=\frac{1}{\sqrt{2}}\{|00,11\rangle+|01,10\rangle\}$\\
        $1$~~~~ & $1$ & $-$ &~~~~~~$|\phi_{7}\rangle=\frac{1}{\sqrt{2}}\{|00,11\rangle-|01,10\rangle\}$\\
        $1$~~~~ & $2$ & $-$ &~~~~~~$|\phi_{8}\rangle=|01,11\rangle$\\
        \hline
        $2$~~~~ & $0$ & $+$ &~~~~~~$|\phi_{13}\rangle=|10,10\rangle$\\
        $2$~~~~ & $1$ & $+$ &~~~~~~$|\phi_{14}\rangle=\frac{1}{\sqrt{2}}\{|10,11\rangle+|11,10\rangle\}$\\
        $2$~~~~ & $1$ & $-$ &~~~~~~$|\phi_{15}\rangle=\frac{1}{\sqrt{2}}\{|10,11\rangle-|11,10\rangle\}$\\
        $2$~~~~ & $2$ & $-$ &~~~~~~$|\phi_{16}\rangle=|11,11\rangle$\\
        \hline
        \hline
    \end{tabular}
\caption{The two-site basis states that appear in Eq.(\ref{eq:7}),
in the form $|w_in_i,w_jn_j\rangle$. The total localized and
conduction electron numbers $w$ and $n$, the eigenvalue $u$ of the
operator $T_{ij}$ defined after Eq.(\ref{eq:8}) are indicated.
$|\phi_{9-12}\rangle$ are respectively obtained from
$|\phi_{5-8}\rangle$ by the action of $T_{ij}$, while the
corresponding Hamiltonian matrix elements are multiplied by the $u$
values of the states.}
    \label{tab:1}
\end{table}

\begin{table}[t!]
    \begin{tabular}{c c c c}
        \hline
        \hline
        $w$~~~~ & ~~~~~$n$~~~~~ & ~~~~~$u$~~~~~ & Three-site basis states\\
        \hline
        $0$~~~~ & $0$ & $+$ & $|\psi_{1}\rangle=|00,00,00\rangle$\\
        $0$~~~~ & $1$ & $+$ & $|\psi_{2}\rangle=\frac{1}{\sqrt{2}}\{|00,00,01\rangle+|01,00,00\rangle\}$\\
        $0$~~~~ & $1$ & $+$ & $|\psi_{3}\rangle=|00,01,00\rangle$\\
        $0$~~~~ & $1$ & $-$ & $|\psi_{4}\rangle=\frac{1}{\sqrt{2}}\{|00,00,01\rangle-|01,00,00\rangle\}$\\
        $0$~~~~ & $2$ & $+$ & $|\psi_{5}\rangle=\frac{1}{\sqrt{2}}\{|00,01,01\rangle-|01,01,00\rangle\}$\\
        $0$~~~~ & $2$ & $-$ & $|\psi_{6}\rangle=|01,00,01\rangle$\\
        $0$~~~~ & $2$ & $-$ & $|\psi_{7}\rangle=\frac{1}{\sqrt{2}}\{|00,01,01\rangle+|01,01,00\rangle\}$\\
        $0$~~~~ & $3$ & $-$ & $|\psi_{8}\rangle=|01,01,01\rangle$\\
        \hline
        $1$~~~~ & $0$ & $+$ & $|\psi_{9}\rangle=|00,00,10\rangle$\\
        $1$~~~~ & $0$ & $+$ & $|\psi_{10}\rangle=|00,10,00\rangle$\\
        $1$~~~~ & $1$ & $+$ & $|\psi_{12}\rangle=\frac{1}{\sqrt{2}}\{|00,00,11\rangle+|01,00,10\rangle\}$\\
        $1$~~~~ & $1$ & $+$ & $|\psi_{13}\rangle=\frac{1}{\sqrt{2}}\{|00,10,01\rangle+|01,10,00\rangle\}$\\
        $1$~~~~ & $1$ & $+$ & $|\psi_{15}\rangle=|00,01,10\rangle$\\
        $1$~~~~ & $1$ & $+$ & $|\psi_{16}\rangle=|00,11,00\rangle$\\
        $1$~~~~ & $1$ & $-$ & $|\psi_{18}\rangle=\frac{1}{\sqrt{2}}\{|00,00,11\rangle-|01,00,10\rangle\}$\\
        $1$~~~~ & $1$ & $-$ & $|\psi_{19}\rangle=\frac{1}{\sqrt{2}}\{|00,10,01\rangle-|01,10,00\rangle\}$\\
        $1$~~~~ & $2$ & $+$ & $|\psi_{21}\rangle=\frac{1}{\sqrt{2}}\{|00,01,11\rangle-|01,01,10\rangle\}$\\
        $1$~~~~ & $2$ & $+$ & $|\psi_{22}\rangle=\frac{1}{\sqrt{2}}\{|00,11,01\rangle-|01,11,00\rangle\}$\\
        $1$~~~~ & $2$ & $-$ & $|\psi_{24}\rangle=|01,00,11\rangle$\\
        $1$~~~~ & $2$ & $-$ & $|\psi_{25}\rangle=|01,10,01\rangle$\\
        $1$~~~~ & $2$ & $-$ & $|\psi_{27}\rangle=\frac{1}{\sqrt{2}}\{|00,01,11\rangle+|01,01,10\rangle\}$\\
        $1$~~~~ & $2$ & $-$ & $|\psi_{28}\rangle=\frac{1}{\sqrt{2}}\{|00,11,01\rangle+|01,11,00\rangle\}$\\
        $1$~~~~ & $3$ & $-$ & $|\psi_{30}\rangle=|01,01,11\rangle$\\
        $1$~~~~ & $3$ & $-$ & $|\psi_{31}\rangle=|01,11,01\rangle$\\
        \hline
        $2$~~~~ & $0$ & $+$ & $|\psi_{33}\rangle=|00,10,10\rangle$\\
        $2$~~~~ & $0$ & $+$ & $|\psi_{34}\rangle=|10,00,10\rangle$\\
        $2$~~~~ & $1$ & $+$ & $|\psi_{36}\rangle=\frac{1}{\sqrt{2}}\{|00,10,11\rangle+|01,10,10\rangle\}$\\
        $2$~~~~ & $1$ & $+$ & $|\psi_{37}\rangle=\frac{1}{\sqrt{2}}\{|10,00,11\rangle+|11,00,10\rangle\}$\\
        $2$~~~~ & $1$ & $+$ & $|\psi_{39}\rangle=|00,11,10\rangle$\\
        $2$~~~~ & $1$ & $+$ & $|\psi_{40}\rangle=|10,01,10\rangle$\\
        $2$~~~~ & $1$ & $-$ & $|\psi_{42}\rangle=\frac{1}{\sqrt{2}}\{|00,10,11\rangle-|01,10,10\rangle\}$\\
        $2$~~~~ & $1$ & $-$ & $|\psi_{43}\rangle=\frac{1}{\sqrt{2}}\{|10,00,11\rangle-|11,00,10\rangle\}$\\
        $2$~~~~ & $2$ & $+$ & $|\psi_{45}\rangle=\frac{1}{\sqrt{2}}\{|00,11,11\rangle-|01,11,10\rangle\}$\\
        $2$~~~~ & $2$ & $+$ & $|\psi_{46}\rangle=\frac{1}{\sqrt{2}}\{|10,01,11\rangle-|11,01,10\rangle\}$\\
        $2$~~~~ & $2$ & $-$ & $|\psi_{48}\rangle=|01,10,11\rangle$\\
        $2$~~~~ & $2$ & $-$ & $|\psi_{49}\rangle=|11,00,11\rangle$\\
        $2$~~~~ & $2$ & $-$ & $|\psi_{51}\rangle=\frac{1}{\sqrt{2}}\{|00,11,11\rangle+|01,11,10\rangle\}$\\
        $2$~~~~ & $2$ & $-$ & $|\psi_{52}\rangle=\frac{1}{\sqrt{2}}\{|10,01,11\rangle+|11,01,10\rangle\}$\\
        $2$~~~~ & $3$ & $-$ & $|\psi_{54}\rangle=|01,11,11\rangle$\\
        $2$~~~~ & $3$ & $-$ & $|\psi_{55}\rangle=|11,01,11\rangle$\\
        \hline
        $3$~~~~ & $0$ & $+$ & $|\psi_{57}\rangle=|10,10,10\rangle$\\
        $3$~~~~ & $1$ & $+$ & $|\psi_{58}\rangle=\frac{1}{\sqrt{2}}\{|10,10,11\rangle+|11,10,10\rangle\}$\\
        $3$~~~~ & $1$ & $+$ & $|\psi_{59}\rangle=|10,11,10\rangle$\\
        $3$~~~~ & $1$ & $-$ & $|\psi_{60}\rangle=\frac{1}{\sqrt{2}}\{|10,10,11\rangle-|11,10,10\rangle\}$\\
        $3$~~~~ & $2$ & $+$ & $|\psi_{61}\rangle=\frac{1}{\sqrt{2}}\{|10,11,11\rangle-|11,11,10\rangle\}$\\
        $3$~~~~ & $2$ & $-$ & $|\psi_{62}\rangle=|11,10,11\rangle$\\
        $3$~~~~ & $2$ & $-$ & $|\psi_{63}\rangle=\frac{1}{\sqrt{2}}\{|10,11,11\rangle+|11,11,10\rangle\}$\\
        $3$~~~~ & $3$ & $-$ & $|\psi_{64}\rangle=|11,11,11\rangle$\\
        \hline
        \hline
    \end{tabular}
\caption{The three-site basis states that appear in Eq.(\ref{eq:7}),
in the form $|w_in_i,w_jn_j,w_kn_k\rangle$. The total localized and
conduction electron numbers $w$ and $n$, the eigenvalue $u$ of the
operator $T_{ik}$ defined after Eq.(\ref{eq:8}) are indicated.
$\{|\psi_{11+3x}\rangle\}$, $x=0,1,\ldots,15$, are respectively
obtained from $\{|\psi_{9+3x}\rangle\}$ by the action of $T_{ik}$,
while the corresponding Hamiltonian matrix elements are multiplied
by the $u$ values of the states.}
    \label{tab:2}
\end{table}

            \begin{table*}[th!]
        \begin{tabular}{c c c c c c c c c c c c c}
            \hline
            \hline
            Phase\TopSpaceInTable & \multicolumn{12}{c}{The interaction constants $K_\alpha$ at the phase sinks}\\
            \cline{2-13}
            sink\TopSpaceInTable\BotSpaceInTable & $t$ & $U$ & $\mu$ & $\nu$ & $J$ & $K$ & $L$ & $P$ & $V_n$ & $V_w$ & $Q$ & $R$ \\
            \hline
            $\delta_{\textrm{dd}}$ & $0$ & $0$ & $-\infty$ & $-\infty$ & $0$ & $0$ & $0$ & $0$ & $0$ & $0$ & $0$ & $0$\\
            $\delta_{\textrm{dD}}$ & $0$ & $\infty$ & $\infty$ & $-\infty$ & $0$ & $0$ & $0$ & $0$ & $0$ & $0$ & $0$ & $0$\\
            $\delta_{\textrm{Dd}}$ & $0$ & $\infty$ & $-\infty$ & $\infty$ & $0$ & $0$ & $0$ & $0$ & $0$ & $0$ & $0$ & $0$\\
            $\delta_{\textrm{DD}}$ & $0$ & $0$ & $\infty$ & $\infty$ & $0$ & $0$ & $0$ & $0$ & $0$ & $0$ & $0$ & $0$\\
            CO$_{\textrm{dd}}$ & $\infty$ & $\infty$ & $-\infty$ & $-\infty$ & $\infty$ & $-\infty$ & $\infty$ & $-\infty$ & $-\infty$ & $\infty$ & $\infty$ & $-\infty$\\
            CO$_{\textrm{dD}}$ & $\infty$ & $\infty$ & $\infty$ & $-\infty$ & $\infty$ & $-\infty$ & $\infty$ & $\infty$ & $-\infty$ & $\infty$ & $\infty$ & $\infty$\\
            CO$_{\textrm{Dd}}$ & $\infty$ & $\infty$ & $-\infty$ & $\infty$ & $\sim50$ & $-\infty$ & $\infty$ & $-\infty$ & $\sim-20$ & $\infty$ & $\infty$ & $-\infty$\\
            CO$_{\textrm{DD}}$ & $\infty$ & $\infty$ & $\infty$ & $\infty$ & $\sim140$ & $-\infty$ & $\infty$ & $\infty$ & $\sim-40$ & $\infty$ & $\infty$ & $\infty$\\
            \hline
            \hline
        \end{tabular}

        \vspace{3mm}

        \begin{tabular}{c c c c c c c c c c c c c}
            \hline
            \hline
            Phase\TopSpaceInTable &  \multicolumn{12}{c}{The runaway coefficients $K_\alpha^\prime/K_\alpha$ at the phase sinks}\\
            \cline{2-13}
            sink\TopSpaceInTable\BotSpaceInTable & $t^\prime/t$ & $U^\prime/U$ & $\mu^\prime/\mu$ & $\nu^\prime/\nu$ & $J^\prime/J$ & $K^\prime/K$ & $L^\prime/L$ & $P^\prime/P$ & $V_n^\prime/V_n$ & $V_w^\prime/V_w$ & $Q^\prime/Q$ & $R^\prime/R$ \\
            \hline
            $\delta_{\textrm{dd}}$, $\delta_{\textrm{DD}}$ & $-$ & $-$ & $4$ & $4$ & $-$ & $-$ & $-$ & $-$ & $-$ & $-$ & $-$ & $-$\\
            $\delta_{\textrm{dD}}$, $\delta_{\textrm{Dd}}$ & $-$ & $4$ & $4$ & $4$ & $-$ & $-$ & $-$ & $-$ & $-$ & $-$ & $-$ & $-$\\
            CO$_{\textrm{dd}}$, CO$_{\textrm{dD}}$ & $2$ & $2$ & $2$ & $4$ & $4/3$ & $2$ & $4/3$ & $2$ & $4/3$ & $2$ & $2$ & $2$\\
            CO$_{\textrm{Dd}}$, CO$_{\textrm{DD}}$ & $2$ & $2$ & $2$ & $4$ & $1$ & $2$ & $4/3$ & $2$ & $1$ & $2$ & $2$ & $2$\\
            \hline
            \hline
        \end{tabular}

        \vspace{3mm}

        \begin{tabular}{c c c c c c c c c c c c c c }
            \hline
            \hline
            Phase\TopSpaceInTable & \multicolumn{12}{c}{The expectation values $M_\alpha$ at the phase sinks} & Character\\
            \cline{2-13}
sink\TopSpaceInTable\BotSpaceInTable & $\langle\hat{t}\rangle$ & $\langle\hat{U}\rangle$ & $\langle\hat{\mu}\rangle$ & $\langle\hat{\nu}\rangle$ & $\langle\hat{J}\rangle$ & $\langle\hat{K}\rangle$ & $\langle\hat{L}\rangle$ & $\langle\hat{P}\rangle$ & $\langle\hat{V_n}\rangle$ & $\langle\hat{V_w}\rangle$ & $\langle\hat{Q}\rangle$ & $\langle\hat{R}\rangle$ & \\
            \hline
            $\delta_{\textrm{dd}}$ & $0$ & $0$ & $0$ & $0$ & $0$ & $0$ & $0$ & $0$ & $0$ & $0$ & $0$ & $0$ & dilute - dilute\\
            $\delta_{\textrm{dD}}$ & $0$ & $0$ & $2$ & $0$ & $1$ & $0$ & $0$ & $0$ & $0$ & $0$ & $0$ & $0$ & dilute - dense\\
            $\delta_{\textrm{Dd}}$ & $0$ & $0$ & $0$ & $2$ & $0$ & $1$ & $0$ & $0$ & $0$ & $0$ & $1$ & $2$ & dense - dilute\\
            $\delta_{\textrm{DD}}$ & $0$ & $2$ & $2$ & $2$ & $1$ & $1$ & $1$ & $2$ & $2$ & $2$ & $-1$ & $-2$ & dense - dense\\
            CO$_{\textrm{dd}}$ & $-a$ & $0$ & $a$ & $0$ & $0$ & $0$ & $0$ & $0$ & $0$ & $0$ & $0$ & $0$ & dilute - charge ord. dilute\\
            CO$_{\textrm{dD}}$ & $-a$ & $0$ & $2-a$ & $0$ & $1-a$ & $0$ & $0$ & $0$ & $0$ & $0$ & $0$ & $0$ & dilute - charge ord. dense\\
            CO$_{\textrm{Dd}}$ & $-a$ & $a$ & $a$ & $2$ & $0$ & $1$ & $0$ & $a$ & $0$ & $a$ & $1$ & $2$ & dense - charge ord. dilute\\
            CO$_{\textrm{DD}}$ & $-a$ & $2-a$ & $2-a$ & $2$ & $1-a$ & $1$ & $1-a$ & $2-a$ & $2-2a$ & $2-a$ & $-1+2a$ & $-2+4a$ & dense - charge ord. dense\\
            \hline
            \hline
\end{tabular}
\caption{Interaction constants $K_\alpha$, runaway coefficients
$K_\alpha^\prime/K_\alpha$, and expectation values
$M_\alpha=\langle\hat{K_\alpha}\rangle$, at the phase sinks. Here,
$\hat{K_\alpha}$ are used as abbreviations for the conjugate
operators for interaction constants $K_\alpha$, \textit{e.g.},
$\langle \hat{t}\rangle=\langle c_i^{\dag}c_j+c_j^{\dag}c_i\rangle$,
$\langle\hat{U}\rangle=\langle n_iw_i+n_jw_j\rangle$, \textit{etc.}
The non-zero hopping expectation value is $-a = -0.629050$.  In the
subscripts in the first columns, the left and right entries refer to
the localized and conduction electrons, respectively, as dilute (d)
or dense (D).} \label{tab:3}
\end{table*}

    \begin{table*}[th!]
        \begin{tabular}{c c c c c c c c c c c c c c c c}
            \hline
            \hline
            Phase\TopSpaceInTable & Boundary & \multicolumn{12}{c}{Interaction constants $K_\alpha$ at the boundary fixed points} & \qquad\qquad\qquad~~~~~~~Relevant & eigenvalue ~~ \\
            \cline{3-14}
            boundary\TopSpaceInTable\BotSpaceInTable & type & $t$ & $U$ & $\mu$ & $\nu$ & $J$ & $K$ & $L$ & $P$ & $V_n$ & $V_w$ & $Q$ & $R$ & & exponent $y_1$\\
            \hline

            CO$_{\textrm{dD}}$/CO$_{\textrm{Dd}}$\TopSpaceInTable & 1st & $\infty$ & $-\infty$ & $\infty$ & $-\infty$ & $\infty$ & $\infty$ & $\infty$ & $-\infty$ & $-\infty$ & $\infty$ & $\infty$ & $-\infty$ & $2\mu-2\nu+J-K$ & 3\\
            \BotSpaceInTable & order &  &  &  &  &  &  &  &  &  &  &  &  & $-Q-2R=0$ & \\

            CO$_{\textrm{dd}}$/$\delta_{\textrm{dd}}$\TopSpaceInTable & 2nd & $\infty$ & $\infty$ & $-\infty$ & $-\infty$ & $\infty$ & $-\infty$ & $\infty$ & $-\infty$ & $-\infty$ & $\infty$ & $\infty$ & $-\infty$ & $t+\mu=1.744253$ & 0.273873\\
            \BotSpaceInTable & order &  &  &  &  &  &  &  &  &  &  &  &  &  & \\

            CO$_{\textrm{dD}}$/$\delta_{\textrm{dD}}$\TopSpaceInTable & 2nd & $\infty$ & $\infty$ & $\infty$ & $-\infty$ & $\infty$ & $-\infty$ & $\infty$ & $-\infty$ & $\infty$ & $\infty$ & $\infty$ & $-\infty$ & $t-\mu-J=1.744253$ & 0.273873\\
            \BotSpaceInTable & order &  &  &  &  &  &  &  &  &  &  &  &  &  & \\

            CO$_{\textrm{Dd}}$/$\delta_{\textrm{Dd}}$\TopSpaceInTable & 2nd & $\infty$ & $\infty$ & $-\infty$ & $\infty$ & $\infty$ & $-\infty$ & $\infty$ & $-\infty$ & $-\infty$ & $\infty$ & $\infty$ & $-\infty$ & $t+U+\mu+P+V_w$ & 0.273873\\
            \BotSpaceInTable & order &  &  &  &  &  &  &  &  &  &  &  &  & $=1.744253$ & \\

            CO$_{\textrm{DD}}$/$\delta_{\textrm{DD}}$\TopSpaceInTable & 2nd & $\infty$ & $\infty$ & $\infty$ & $\infty$ & $\infty$ & $-\infty$ & $\infty$ & $\infty$ & $-\infty$ & $\infty$ & $\infty$ & $\infty$ & $t-U-\mu-J-L-P$ & 0.273873\\
            \BotSpaceInTable & order &  &  &  &  &  &  &  &  &  &  &  &  & $-2V_n-V_w+2Q+4R=0$ & \\

            CO$_{\textrm{dd}}$/CO$_{\textrm{dD}}$\TopSpaceInTable & 2nd & $\infty$ & $\infty$ & $-\infty$ & $-\infty$ & $\infty$ & $-\infty$ & $\infty$ & $-\infty$ & $-\infty$ & $\infty$ & $\infty$ & $-\infty$ & $2\mu+J=0$ & 1.420396\\
            \BotSpaceInTable & order &  &  &  &  &  &  &  &  &  &  &  &  &  & \\

            CO$_{\textrm{Dd}}$/CO$_{\textrm{DD}}$\TopSpaceInTable & 2nd & $\infty$ & $\infty$ & $-\infty$ & $\infty$ & $\infty$ & $-\infty$ & $\infty$ & $-\infty$ & $\infty$ & $\infty$ & $\infty$ & $-\infty$ & $2U+2\mu+J+L+2P$ & 1.420396\\
            \BotSpaceInTable & order &  &  &  &  &  &  &  &  &  &  &  &  & $+2V_n+2V_w-2Q-4R=0$ & \\

            \hline
            \hline
        \end{tabular}
\caption{Interaction constants $K_\alpha$ and relevant eigenvalue
exponents $y_1$ at the phase boundary fixed points.  For first-order
phase transitions, $y_1=d=3$.}
        \label{tab:4}
    \end{table*}

    With these basis states, Eq.(\ref{eq:6}) can be rewritten as
    \begin{multline}
        \label{eq:7}
        \langle\phi _{p}|e^{-\beta^{\prime }\mathcal{H}^{\prime}_{i,k}}|\phi_{\bar{p}}\rangle=
        \sum_{\substack{u,v,\\\bar{u},\bar{v},s}}\sum_{\substack{q,\bar{q}}}\langle\phi_p|u_iv_k\rangle\langle u_is_jv_k|\psi_q\rangle \cdot\\
        \langle\psi_q|e^{-\beta\mathcal{H}_{i,j}-\beta\mathcal{H}_{j,k}}|\psi_{\bar{q}}\rangle
        \langle\psi_{\bar{q}}|\bar{u}_is_j\bar{v}_k\rangle\langle\bar{u}_i\bar{v}_k|\phi_{\bar{p}}\rangle.
    \end{multline}
\noindent Once written in the basis states $\{|\phi_{p}\rangle\}$,
the block-diagonal renormalized matrix has 13 independent elements,
which means that renormalization-group transformation of the
Hamiltonian generates 9 more interaction constants apart from $t$,
$U$, $\mu$, and $\nu$. In this 13-dimensional interaction space, the
form of the Hamiltonian stays closed under renormalization-group
transformations. This Hamiltonian is
    \begin{equation}
        \label{eq:8}
        \begin{split}
            -\beta\mathcal{H}_{i,j}&=t\left(c_i^{\dag}c_j+c_j^{\dag}c_i\right)+U\left(n_iw_i+n_jw_j\right)\\
            &+\mu\left(n_i+n_j\right)+\nu\left(w_i+w_j\right)+Jn_in_j\\
            &+Kw_iw_j+Ln_in_jw_iw_j+P\left(n_iw_j+n_jw_i\right)\\
            &+V_nn_in_j(w_i+w_j)+V_w(n_i+n_j)w_iw_j\\
            &+QT_{ij}w_iw_j+RT_{ij}(w_i+w_j)+G,
        \end{split}
    \end{equation}
where $T_{ij}$ is a local operator that switches the conduction
electron states of sites $i$ and $j$:
$T_{ij}|w_in_i,w_jn_j\rangle=u|w_in_j,w_jn_i\rangle$ with $u=1$ for
$n_i+n_j<2$ and $u=-1$ otherwise. When $T_{ik}$ is applied, further
below, to three consecutive sites $i$, $j$, $k$,
$T_{ik}|w_in_i,w_jn_j,w_kn_k\rangle=u|w_in_k,w_jn_j,w_kn_i\rangle$
with $u=1$ for $n_i+n_j+n_k<2$ and $u=-1$ otherwise.

\begin{figure}[h!]
\centering
\includegraphics*[scale=1.0]{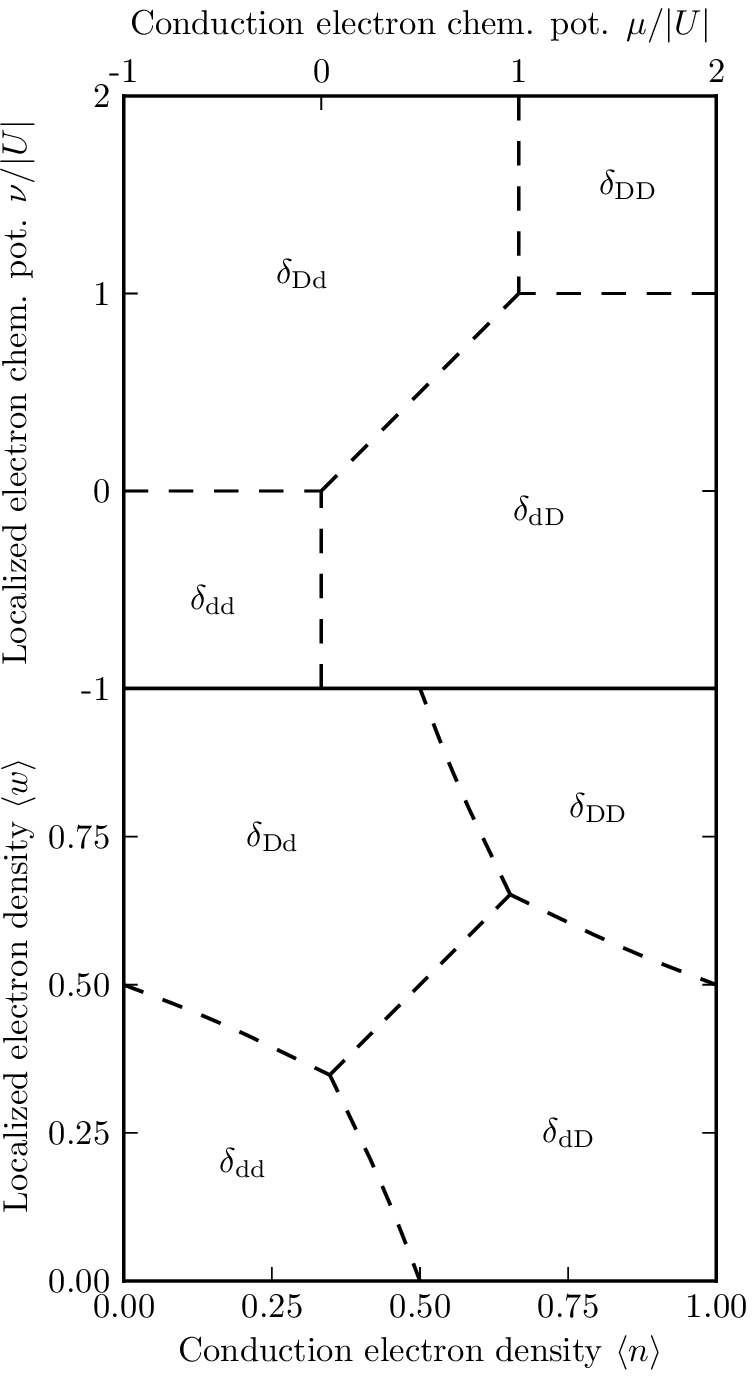}
\caption{Renormalization-group flow basins of the $t=0$ classical
submodel, in the chemical potentials (upper panel) and densities
(lower panel) of the localized and conduction electrons.  In phase
subscripts throughout this paper, the first and second subscripts
respectively describe localized and conduction electron densities,
as dilute (d) or dense (D).  The dashed lines are not phase
transitions, but smooth changes between the four different density
regions of the disordered ($\delta$) phase.} \label{fig:1}
\end{figure}

To extract the renormalization-group recursion relations, we
consider the matrix elements
$\gamma_{p,\bar{p}}\equiv\langle\phi_{p}|e^{-\beta^\prime\mathcal{H}^\prime_{i,k}}|\phi_{\bar{p}}\rangle$.
With $\gamma_{9,9}=\gamma_{5,5}$, $\gamma_{10,10}=\gamma_{6,6}$,
$\gamma_{11,11}=\gamma_{7,7}$, and $\gamma_{12,12}=\gamma_{8,8}$, 12
out of 16 diagonal elements are independent and, with
$\gamma_{10,11}=\gamma_{11,10}=-\gamma_{7,6}=-\gamma_{6,7}$, only
one of the 4 off-diagonal elements is independent, summing up to 13
independent matrix elements. Thus we obtain the renormalized
interaction constants in terms of $\{\gamma\}$, defining
$\gamma_p\equiv\gamma_{p,p}$ for the diagonal elements and
$\gamma_0\equiv\gamma_{6,7}$ for the only independent off-diagonal
element:
    \begin{multline}
        \label{eq:9}
        t^\prime=\frac{1}{2}\ln\frac{\gamma_2}{\gamma_3},\quad
        U^\prime=\ln\frac{\gamma_1\gamma_6\gamma_0}{\gamma_2\gamma_5},\quad
        \mu^\prime=\frac{1}{2}\ln\frac{\gamma_2\gamma_3}{\gamma_1^2},\\
        \nu^\prime=\frac{1}{2}\ln\frac{\gamma_2\gamma_5^2\gamma_7}{\gamma_1^2\gamma_3\gamma_6},\quad
        J^\prime=\ln\frac{\gamma_1\gamma_4}{\gamma_2\gamma_3},\\
        K^\prime=\frac{1}{2}\ln\frac{\gamma_1^2\gamma_3\gamma_6^2\gamma_{13}^2\gamma_{15}}{\gamma_2\gamma_5^4\gamma_7^2\gamma_{14}},\quad
        L^\prime=\ln\frac{\gamma_1\gamma_4\gamma_6^2\gamma_7^2\gamma_{13}\gamma_{16}}{\gamma_2\gamma_3\gamma_5^2\gamma_8^2\gamma_{14}\gamma_{15}},\\
        P^\prime=\ln\frac{\gamma_1\gamma_6}{\gamma_2\gamma_5\gamma_0},\quad
        V_n^\prime=\ln\frac{\gamma_2\gamma_3\gamma_5\gamma_8}{\gamma_1\gamma_4\gamma_6\gamma_7},\quad
        V_w^\prime=\ln\frac{\gamma_2\gamma_5^2\gamma_{14}}{\gamma_1\gamma_6^2\gamma_{13}},\\
        Q^\prime=\frac{1}{2}\ln\frac{\gamma_2\gamma_7^2\gamma_{14}}{\gamma_3\gamma_6^2\gamma_{15}},\quad
        R^\prime=\frac{1}{2}\ln\frac{\gamma_3\gamma_6}{\gamma_2\gamma_7},\quad
        G^\prime=\ln\gamma_1.
    \end{multline}
The matrix elements $\{\gamma\}$ of the exponentiated renormalized
Hamiltonian are connected, by Eq.(\ref{eq:7}), to the matrix
elements,
$\eta_{q,\bar{q}}\equiv\langle\psi_{q}|e^{-\beta\mathcal{H}_{i,j}-\beta\mathcal{H}_{j,k}}|\psi_{\bar{q}}\rangle$
of the exponentiated unrenormalized Hamiltonian,
\begin{equation}
      \begin{split}
            \gamma_0=&\eta_{12,18}+\eta_{21,27}+\eta_{36,42}+\eta_{45,51},\\
            \gamma_1=&\eta_{1}+\eta_{3}+\eta_{10}+\eta_{16},\\
            \gamma_2=&\eta_{3}+\eta_{7}+\eta_{13}+\eta_{28},\\
            \gamma_3=&\eta_{4}+\eta_{5}+\eta_{19}+\eta_{22},\\
            \gamma_4=&\eta_{6}+\eta_{8}+\eta_{25}+\eta_{31},\\
            \gamma_5=&\eta_{9}+\eta_{15}+\eta_{33}+\eta_{39},\\
            \gamma_6=&\eta_{12}+\eta_{27}+\eta_{36}+\eta_{51},\\
            \gamma_7=&\eta_{18}+\eta_{21}+\eta_{42}+\eta_{45},\\
            \gamma_8=&\eta_{24}+\eta_{30}+\eta_{48}+\eta_{54},\\
            \gamma_{13}=&\eta_{34}+\eta_{40}+\eta_{57}+\eta_{59},\\
            \gamma_{14}=&\eta_{37}+\eta_{52}+\eta_{58}+\eta_{63},\\
            \gamma_{15}=&\eta_{43}+\eta_{46}+\eta_{60}+\eta_{61},\\
            \gamma_{16}=&\eta_{49}+\eta_{55}+\eta_{62}+\eta_{64}.
            \label{eq:10}
      \end{split}
\end{equation}
The matrix elements $\eta_{q,\bar{q}}$ can be obtained in
terms of the unrenormalized interactions via exponentiating
the unrenormalized Hamiltonian matrix whose elements are
given in Appendix B.

\begin{figure*}[th!]
\centering
\includegraphics*[scale=1.0]{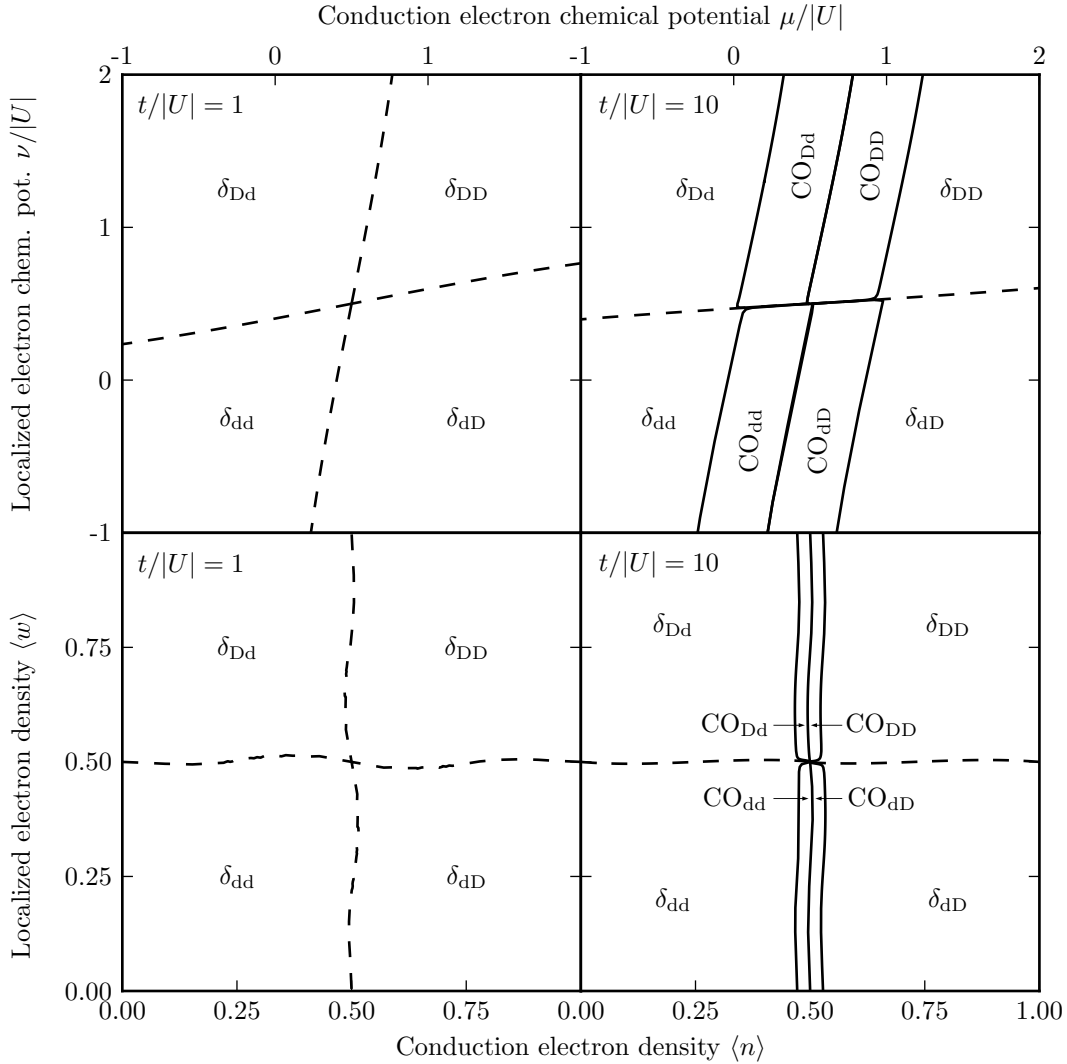}
\caption{Constant $t/|U|$ cross-sections of the phase diagram for
interaction $|U|=0.1$, in terms of the chemical potentials (upper
panels) and densities (lower panels) of the localized and conduction
electrons. In phase subscripts throughout this paper, the first and
second subscripts respectively describe localized and conduction
electron densities, as dilute (d) or dense (D).  The full lines are
second-order phase transitions. The dashed lines are not phase
transitions, but smooth changes between the different density
regions of the disordered ($\delta$) phase. The charge-ordered
phases are denoted by CO. Details are shown in Fig. 4. Thus, for low
values of the interaction, all phase boundaries are second order and
there is no phase coexistence.} \label{fig:2}
\end{figure*}

\begin{figure}[th!]
\centering
\includegraphics*[scale=1]{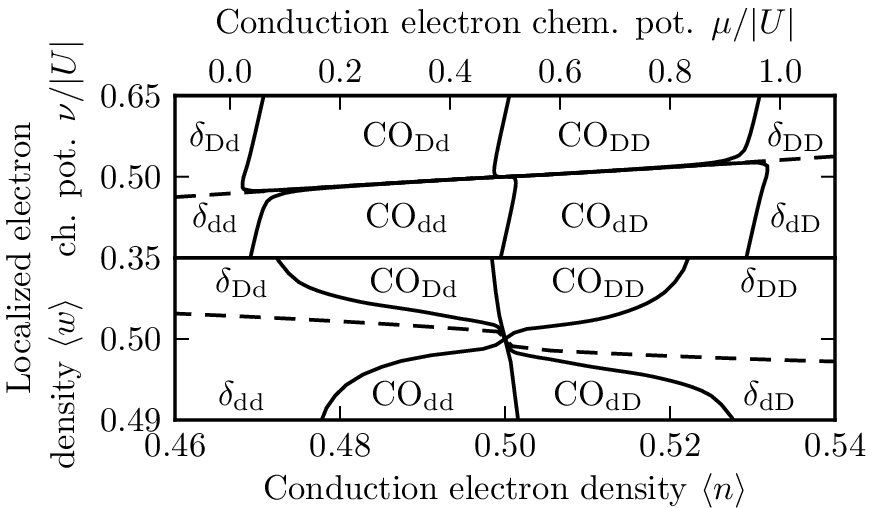}
\caption{Zoomed portion of Fig. 3, for the $|U|=0.1$, $t/|U|=10$
phase diagram.} \label{fig:3}
\end{figure}

    \subsection{Renormalization-Group Transformation in $d>1$}

Equations (\ref{eq:9}) and (\ref{eq:10}), together with Appendix B,
constitute the renormalization-group recursion relations for $d=1$,
in the form $\vec{K}^\prime=R(\vec{K})$, where $\vec{K}=(t, U, \mu,
\nu, J, K, L, P, V_n, V_w, Q, R, G)$.  To generalize to higher
dimension $d>1$, we use the Migdal-Kadanoff procedure \cite{Migdal,
Kadanoff},
\begin{equation}
        \label{eq:11}
        \vec{K}^\prime=b^{d-1}R(\vec{K}),
\end{equation}
where $b=2$ is the rescaling factor and $R$ is the
renormalization-group transformation in $d=1$ for the interaction
constants vector $\vec{K}$. This procedure is exact for
$d$-dimensional hierarchical lattices
\cite{BerkerOslund79,Kaufman81,Kaufman84} and a very good
approximation for hypercubic lattices for obtaining complex phase
diagrams.

Each phase in the phase diagram has its own (stable) fixed point(s),
which is called a phase sink (Table \ref{tab:3}). All points within
a phase flow to the sink(s) of that phase under successive
renormalization-group transformations. Phase boundaries also have
their own (unstable) fixed points (Table \ref{tab:4}), where the
relevant exponent analysis gives the order of the phase transition.
Thus, the repartition of the renormalization-group flows determine
the phase diagram in thermodynamic-field space.  Matrix
multiplications, along the renormalization-group trajectory, with
the derivative matrix of the recursion relations relate the
expectation values at the starting point of the trajectory to the
expectations values at the phase sink.  The latter are determined
(Table III) by the left eigenvector, with eigenvalue $b^d$, of the
recursion matrix at the sink, where $b=2$ is the length-rescaling
factor of the renormalization-group transformation.  When the
expectation values are thus calculated for the points of the phase
boundary, the phase diagram in density space is
determined.\cite{McKay84, Hinczewski06b}

    \section{Global Phase Diagram of SFKM}

The global phase diagram of SFKM is calculated, as described above,
for the whole range of the interactions ($t$, $U$, $\nu$, $\mu$).
The global phase diagram is thus 4-dimensional.  $1/t$ can be taken
as the temperature variable.  We present the calculated global phase
diagram in four subsections: The first subsection gives the $t=0$
classical submodel. The other subsections are devoted to small,
intermediate, and large values of the interaction $|U|$. We present
constant $t/|U|$ cross sections in terms of the localized and
conduction electron chemical potentials $\nu/|U|$ and $\mu/|U|$and
in terms of the localized and conduction electron densities $\langle
w_i\rangle$ and $\langle n_i\rangle$.

\begin{figure*}[th!]
\centering
\includegraphics*[scale=1.0]{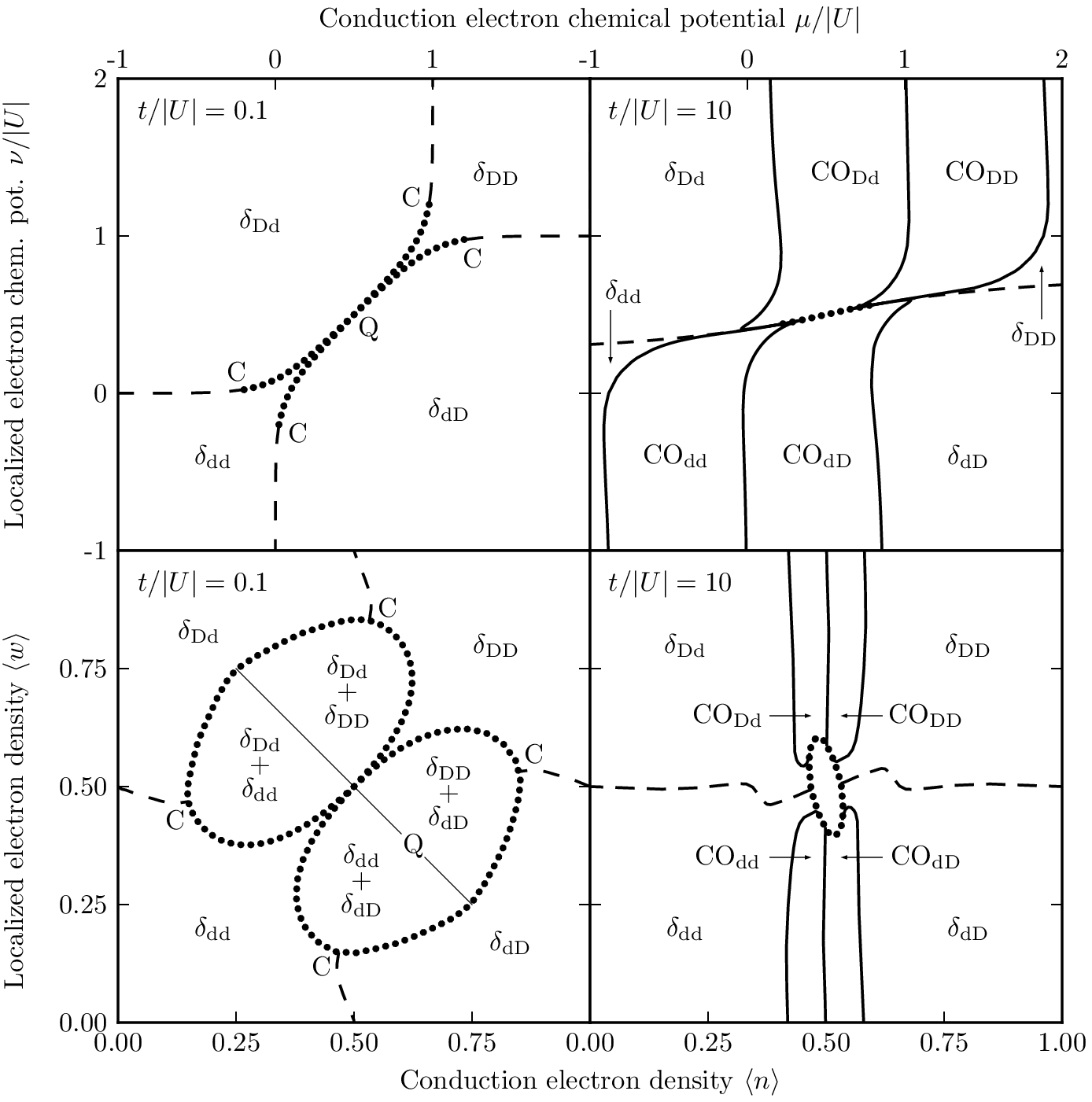}
\caption{Constant $t/|U|$ cross-sections of the phase diagram for
interaction $|U|=1$, in terms of the chemical potentials (upper
panels) and densities (lower panels) of the localized and conduction
electrons. The dotted and thick full lines are respectively first-
and second-order phase transitions. Phase separation, \textit{i.e.},
phase coexistence occurs inside the dotted boundaries, as identified
in the figure. The details of the coexistence region in the
lower-right panel are given in Fig. 6.  The quadruple point Q tie
line is shown as the thin straight line. The dashed lines are not
phase transitions, but smooth changes between the different density
regions of the disordered ($\delta$) phase.} \label{fig:4}
\end{figure*}

\begin{figure}[th!]
\centering
\includegraphics*[scale=1]{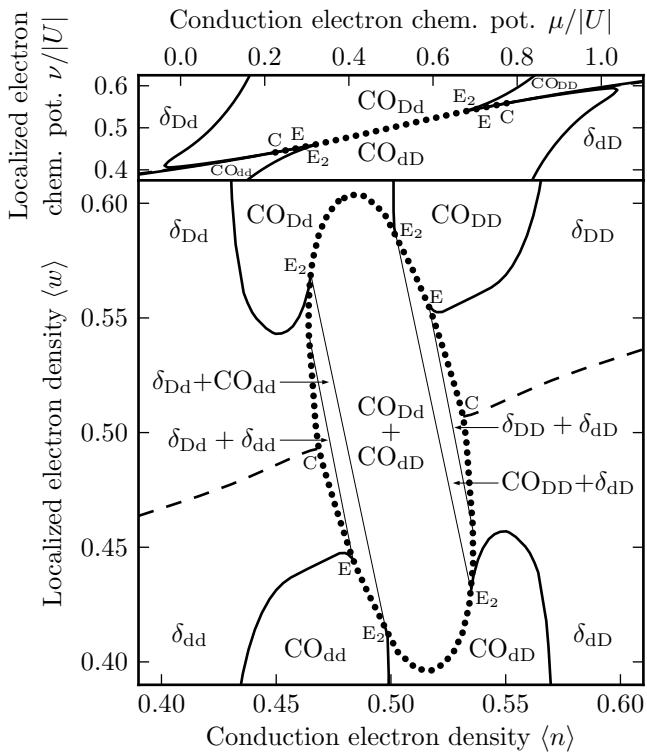}
\caption{Zoomed portion of Fig. 5, for the $|U|=1$, $t/|U|=10$ phase
diagram.  The coexistence tie lines of the critical endpoints E and
of the double critical endpoints E$_2$ are shown.  Inside each
region delimited by dotted lines and the endpoint tie lines, phase
separation, \textit{i.e.}, phase coexistence occurs between phases
as identified on this figure.} \label{fig:5}
\end{figure}

\begin{figure*}[th!]
\centering
\includegraphics*[scale=1]{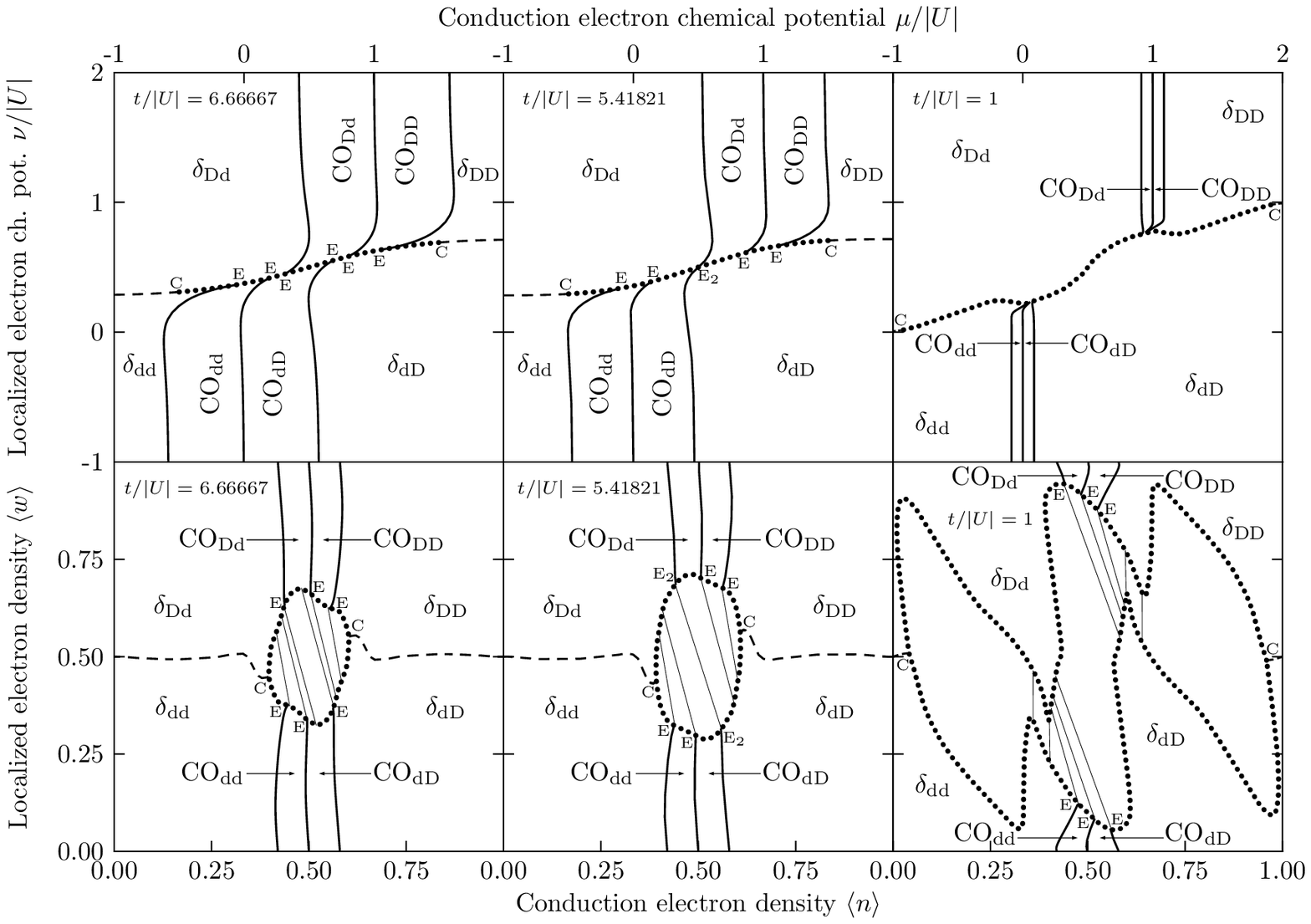}
\caption{Constant $t/|U|$ cross-sections of the phase diagram for
interactions $|U|= 1.5, 1.845628, 10$ (left to right), in terms of
the chemical potentials (upper panels) and densities (lower panels)
of the localized and conduction electrons. The dotted and thick full
lines are respectively first- and second-order phase transitions.
The tie lines of the critical endpoints E and of the double critical
endpoints E$_2$ are shown by thin straight lines.  Phase separation,
\textit{i.e.}, phase coexistence occurs within the regions bounded
by the dotted lines and these endpoint tie lines, this coexistence
being between the phases seen on each side of the dotted lines. In
the upper-right chemical-potential panel, the first-order phase
boundary exhibits, from the left to right, a sequence of the
maximum, minimum, maximum, minimum points; the four corresponding
tie lines are also shown in the lower-right density panel.  These
tie lines abut, on one end, very near maxima and minima of the lower
and upper branches of the coexistence boundaries.  The dashed lines
are not phase transitions, but smooth changes between the different
density regions of the $\delta$ phase.} \label{fig:5}
\end{figure*}

\subsection{The Classical Submodel $t=0$}

Setting the quantum effect to zero, $t=0$, yields the classical
submodel, closed under the renormalization-group flows. The global
flow basins in $\nu/|U|$ and $\mu/|U|$ are the same for all $U$,
given in Fig. \ref{fig:1}.  There exist four regions of a disordered
phase within this submodel, which are
localized-dilute-conduction-dilute,
localized-dilute-conduction-dense,
localized-dense-conduction-dilute, and
localized-dense-conduction-dense regions, denoted by
$\delta_{\textrm{dd}}$, $\delta_{\textrm{dD}}$,
$\delta_{\textrm{Dd}}$, and $\delta_{\textrm{DD}}$. [In phase
subscripts throughout this paper, the first and second subscripts
respectively describe localized and conduction electron densities,
as dilute (d) or dense (D).] In the renormalization-group flows,
each $\delta$ region is the basin of attraction of its own sink. The
dashed lines between the different regions are not phase boundaries,
but smooth transitions (such as the supercritical liquid - gas or
up-magnetized - down-magnetized transitions), which are controlled
by zero-coupling null fixed points.\cite{Berker0}

It should be noted that the Suzuki-Takano and Migdal-Kadanoff
methods are actually exact for this classical submodel, and yield
exactly the same picture as obtained in \cite{Datta99}.

\subsection{The Small $|U|$ Regime}

\begin{figure}[th!]
\centering
\includegraphics*[scale=1]{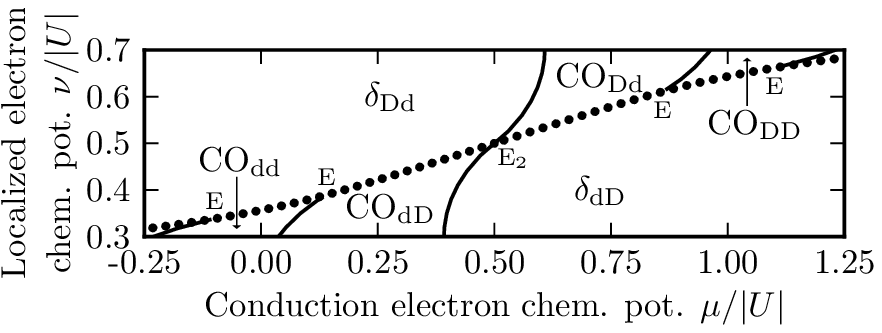}
\caption{Zoomed portions of Fig. 7, for the $|U| = 1.845628, t/|U| =
5.41821$ phase diagram.} \label{fig:6}
\end{figure}

\begin{figure}[th!]
\centering
\includegraphics*[scale=1]{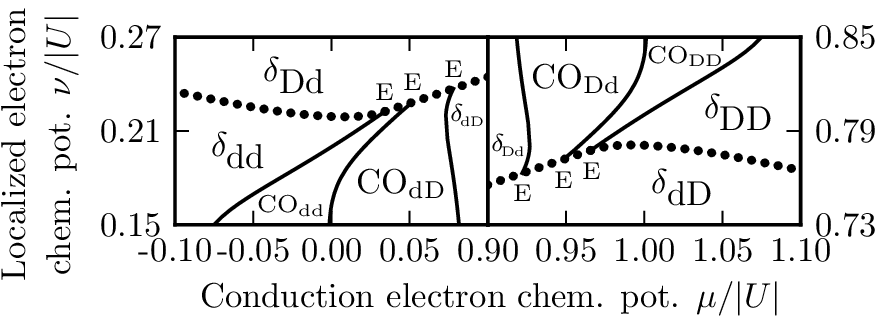}
\caption{Two zoomed portions of Fig. 7, for the $|U|=10, t/|U|=1$
phase diagram.} \label{fig:7}
\end{figure}

In this subsection, we present our results for $|U|=0.1$,
representative of the weak-interaction regime.  The $t=0$ phase
diagram of Fig. 2 evolves under the introduction of quantum effects
via a non-zero hopping strength $t$. It should be noted that
increasing the dimensionless Hamiltonian parameter $t$ is equivalent
to reducing temperature, as in all renormalization-group studies.
The first effect is the decrease and elimination (left panels of
Fig. 3) of the (smooth) passage between the $\delta_{\textrm{Dd}}$
and $\delta_{\textrm{dD}}$ regions. With this elimination, all four
regions meet at $\nu/|U|=\mu/|U|=0.5$ and $\langle
w_i\rangle=\langle n_i\rangle=0.5$, the half filling of both
localized and conduction electrons.  With increasing $t$ (equivalent
to decreasing temperature), four new, charge-ordered (CO) phases
emerge at $t\simeq0.6$.  The CO phases occur at and near half
filling of conduction electrons for the entire range of localized
electron densities.  The CO phases grow with increasing $t$
(decreasing temperature) until saturation at high $t$ (right panels
of Fig. 3).

All of the new CO phases have non-zero hopping density $\langle
c_i^{\dag}c_j+c_j^{\dag}c_i\rangle=-a=-0.629050$ at their phase
sinks. The expectation values at the sinks are evaluated as the left
eigenvector of the recursion matrix with eigenvalue $b^d$
\cite{McKay84}. In the CO phases, the hopping strength $t$ diverges
to infinity under repeated renormalization-group transformations
(whereas in the $\delta$ phases, $t$ vanishes under repeated
renormalization-group transformations).  The localized electron
density is $\langle w_i+w_j\rangle=0$ at the sinks of
CO$_{\textrm{dd}}$ and CO$_{\textrm{dD}}$, while $\langle
w_i+w_j\rangle=2$ at the sinks of CO$_{\textrm{Dd}}$ and
CO$_{\textrm{DD}}$, which throughout the corresponding phases
calculationally translates \cite{McKay84} as low ($d$) and high
($D$) localized electron densities, respectively.  Recall that on
phase labels (CO and $\delta$) throughout this paper, the first and
second subscripts respectively describe localized and conduction
electron densities.

The conduction electron density is $\langle n_i+n_j\rangle=
a=0.629050$ at the sinks of CO$_{\textrm{dd}}$ and
CO$_{\textrm{Dd}}$, while $\langle n_i+n_j\rangle=2-a=1.370950$ at
the sinks of CO$_{\textrm{dD}}$ and CO$_{\textrm{DD}}$.  The
nearest-neighbor conduction electron number correlation is $\langle
n_in_j\rangle=0$ at the sinks of CO$_{\textrm{dd}}$ and
CO$_{\textrm{Dd}}$, while $\langle n_in_j\rangle=1-a=0.370950$ at
the sinks of CO$_{\textrm{dD}}$ and CO$_{\textrm{DD}}$.
Consequently, for conduction electrons, if a given site is occupied,
its nearest-neighbor site is empty at the sinks of
CO$_{\textrm{dd}}$ and CO$_{\textrm{Dd}}$. The CO$_{\textrm{dD}}$
and CO$_{\textrm{DD}}$ phases are connected to the
CO$_{\textrm{dd}}$ and CO$_{\textrm{Dd}}$ phases by particle-hole
interchange on the conduction electrons. Thus, in the CO phases, the
lattice can be divided into two sublattices with different electron
densities. The behavior at the CO sinks therefore indicates charge
ordered phases at finite temperatures, as also previously seen in
ground-state studies \cite{BrandtSchmidt87,GruberIwanski,Stasyuk}.
Note that this charge ordering is a purely quantum mechanical effect
caused by hopping, since the SFKM Hamiltonian [Eq.(\ref{eq:1})]
studied here does not contain an interaction between electrons at
different sites.

In the small $|U|$ regime, all phase boundaries around the CO phases
are second order.  As seen in the expanded Fig. 4, all four CO
phases and all four regions of the $\delta$ phase (as narrow
slivers) meet at $\nu/|U|=\mu/|U|=0.5$ and $\langle
w_i\rangle=\langle n_i\rangle=0.5$, half-filling point of both
localized and conduction electrons. All characteristics of the sinks
and boundary fixed points are given in Tables \ref{tab:3} and
\ref{tab:4}.

    \subsection{The Intermediate $|U|$ Regime}

In this subsection, the phase diagram for $|U|=1$, representative of
the intermediate-interaction regime, is presented. Fig. \ref{fig:4}
gives constant $t/|U|$ cross-sections.  First-order phase boundaries
appear in the central region of the phase diagram, at and near the
half filling of both localized and conduction electrons.

For low values of $t$ (left panels of Fig. 5), equivalent to high
temperatures, two first-order phase boundaries, bounded by four
critical points C, pinch at a quadruple point Q. In the (left-lower)
density-density phase diagram, four phase separation (coexistence)
regions mark the first-order phase transitions. Inside these
regions, coexistence (phase separation) occurs between the phases on
each side of these regions, as indicated on the figure. The tie line
of the quadruple point is shown as a thin straight line. All four
$\delta$ phases coexist (phase separate) on this line.

As $t$ increases (temperature decreases), the four charge-ordered CO
phases appear again at $t\simeq0.6$, as seen in the leftmost panels
of Fig. 1. The CO phases again occur at and near half filling of
conduction electrons for the entire range of localized electron
densities.  In the right panels of Fig. 5, the second-order
transition lines bounding the CO phases terminate at two critical
endpoints E \cite{Berker0} and two double critical endpoints E$_2$
on the first-order line in the central region (zoomed in Fig. 6).
Thus, first-order transitions and phase separation occur between the
pairs of $\delta_{\textrm{Dd}}$ and
$\delta_{\textrm{dd}}$,$\delta_{\textrm{Dd}}$ and
CO$_{\textrm{dd}}$,CO$_{\textrm{Dd}}$ and
CO$_{\textrm{dD}}$,CO$_{\textrm{DD}}$ and $\delta_{\textrm{dD}}$,
$\delta_{\textrm{DD}}$ and $\delta_{\textrm{dD}}$ phases, as
indicated on Fig. 6, at and near the half filling of both localized
and conduction electrons.

The evolution of the phase diagrams between right and left panels of
Fig. 5 are shown in Fig. 1.

    \subsection{The Large $|U|$ Regime}

The evolution of the global phase diagram, as the interaction
strength is increased, is seen in the phase diagrams in Fig. 7.  The
CO phases emerge again at $t\simeq0.6$.  With increasing $t$
(decreasing temperature), the CO phases grow, until saturation seen
in Fig. 7. The topology of the phase diagram with five phases stays
the same for all $t\gtrsim0.6$.

The constant $t/|U|$ cross-sections of the phase diagram are given
in Fig. 7. For $U=1.5$, the double critical endpoints E$_2$ have
split into pairs of simple critical endpoints E, resulting in six
separate critical endpoints.  For $U=1.845628$, the inner two
critical endpoints have merged into a single double critical
endpoint.  For $U=10$, the double critical endpoint has split into
two critical endpoints and the critical lines in the low-density and
high-density localized electrons regions have disconnected from each
other.  In this strong interaction limit, the homogenous
(non-phase-separated) charge-ordered phases occur again at and near
half filling of conduction electrons, but at the low- or
high-density limit of the localized electrons.  Away from these
limits, the charge-ordered phases occur in coexistence
(phase-separated from) the disordered phases.  At and near the half
filling of both localized and conduction electrons, the coexistence
of the disordered phases $\delta_{\textrm{Dd}}$ and
$\delta_{\textrm{dD}}$ occurs.  Two sets of three critical lines
terminate in separate endpoints, as seen in the zoomed Fig. 9. In
this case, a characteristic shape of the density phase diagrams,
which we dub \emph{chimaera} coexistence, emerges.  In the
\emph{chimaera} phase diagram, coexistence can be found for
essentially the entire range of conduction electrons densities or
for most of the range of localized electron densities.  In the
upper-right chemical-potential panel of Fig. 7, the first-order
phase boundary exhibits, from the left to right, a sequence of the
maximum, minimum, maximum, minimum points; the four corresponding
tie lines are also shown in the lower-right density panel.  These
tie lines abut, on one end, very near maxima and minima of the lower
and upper branches of the coexistence boundaries, thereby
underpinning the distinctive \emph{chimaera} topology.

\section{Conclusion}

With this research, we have obtained the global phase diagram of the
$d=3$ SFKM, which exhibits a fairly rich collection of phase diagram
topologies:

For the $t=0$ classical submodel, we have obtained disordered
($\delta$) regions, dilute and dense separately for localized and
conduction electrons, but no phase transition between them.  The
repartition of these regions, delimited by renormalization-group
flows, quantitatively stays the same for the whole $|U|$ range and
is exactly as obtained in Ref. \cite{Datta99}.  For the whole $|U|$
range and $0<t\lesssim0.6$, the classical submodel phase diagram is
perturbed in such a way that regions $\delta_{\textrm{dd}}$ and
$\delta_{\textrm{DD}}$ intercede between regions
$\delta_{\textrm{dD}}$ and $\delta_{\textrm{Dd}}$, resulting in the
shrinking and disappearing of the $\delta_{\textrm{dD}}$ to
$\delta_{\textrm{Dd}}$ passage.

All $\delta$ regions have vanishing hopping density at their
corresponding sinks. For the whole $|U|$ range, upon increasing $t$
(lowering temperature), at $t\simeq0.6$ four new phases (CO) emerge
with non-zero hopping density of $-a = -0.629050$ at their sinks.
These CO phases are also either dilute or dense, separately, in the
localized and conduction electrons (CO$_{\textrm{dd}}$,
CO$_{\textrm{dD}}$, CO$_{\textrm{Dd}}$, and CO$_{\textrm{DD}}$) and
are all charge ordered in the conduction electrons, a wholly quantum
mechanical effect. In these CO phases the bipartite lattice is
divided into two sublattices of alternating electron density.  The
CO phases occur at or near the half filling of conduction electrons.
The phase diagrams with all five phases for $t\gtrsim0.6$ exhibit
different topologies, for the small, intermediate, and large $|U|$
regimes:

For the small $|U|$ (weak-interaction) regime, all phase boundaries
are second order. All five phases meet at $\nu/|U|=\mu/|U|=0.5$ and
$\langle w_i\rangle=\langle n_i\rangle=0.5$, the half-filling point
of both localized and conduction electrons.

For the intermediate $|U|$ (intermediate-interaction) regime, a
first-order phase boundary emerges in the central region of the
phase diagram. This first-order boundary is centered at
$\nu/|U|=\mu/|U|=0.5$ and is bounded by two critical points C. The
second-order lines bounding the CO phases terminate at critical
endpoints E and double critical endpoints E$_2$ on the first-order
boundary.  Due to this first-order phase transition at and near the
half filling of both localized and conduction electrons, a rich
variety of phase separation (phase coexistence) occurs, as indicated
on Figs. 1,5,6,7.

For the large $|U|$ (strong-interaction) regime, as $|U|$ is
increased, the critical endpoints pass through each other by merging
and unmerging as double critical endpoints.  For large $|U|$, the
CO$_\textrm{Dd}$ and CO$_\textrm{DD}$ phases are detached from the
CO$_\textrm{dd}$ and CO$_\textrm{dD}$ phases, forming two separate
bundles, at high- and low-densities of localized electrons
respectively. First-order transitions occur between the variously
dense and dilute $\delta$. The global phase diagram underpinning all
of these cross-sections is decidedly quite complex.

\begin{acknowledgments}
We thank J. L. Lebowitz for suggesting this problem to us.  Support
by the Alexander von Humboldt Foundation, the Scientific and
Technological Research Council of Turkey (T\"UB\.ITAK), and the
Academy of Sciences of Turkey is gratefully acknowledged.  O.S.S.
gratefully acknowledges support from the Scientist Supporting
Office T\"UB\.ITAK-B\.IDEB.
\end{acknowledgments}

\appendix

\section*{APPENDIX A: BLOCK-DIAGONAL RENORMALIZED HAMILTONIAN}

The matrix elements of the block-diagonal renormalized 2-site
Hamiltonian in the $\{|\phi_p\rangle\}$ basis are given in
Eq.(\ref{eq:A1}), where
$\langle\phi_p|-\beta^\prime\mathcal{H}^\prime_{i,k}|\phi_p\rangle=\epsilon_p+G^\prime$
for the 12 independent diagonal elements and
$\langle\phi_6|-\beta^\prime\mathcal{H}^\prime_{i,k}|\phi_7\rangle=\epsilon_0$
for the only independent off-diagonal element:
\begin{multline}
        \label{eq:A1}
        \epsilon_1=0,\quad
        \epsilon_2=t^\prime+\mu^\prime,\quad
        \epsilon_3=-t^\prime+\mu^\prime,\quad
        \epsilon_4=2\mu^\prime+J^\prime,\\
        \epsilon_5=\nu^\prime+R^\prime,\quad
        \epsilon_6=t^\prime+U^\prime/2+\mu^\prime+\nu^\prime+P^\prime/2+R^\prime,\\
        \epsilon_7=-t^\prime+U^\prime/2+\mu^\prime+\nu^\prime+P^\prime/2-R^\prime,\\
        \epsilon_8=U^\prime+2\mu^\prime+\nu^\prime+J^\prime+P^\prime+V_n^\prime-R^\prime,\\
        \epsilon_{13}=2\nu^\prime+K^\prime+Q^\prime+2R^\prime,\\
        \epsilon_{14}=t^\prime+U^\prime+\mu^\prime+2\nu^\prime+K^\prime+P^\prime+V_w^\prime+Q^\prime+2R^\prime,\\
        \epsilon_{15}=-t^\prime+U^\prime+\mu^\prime+2\nu^\prime+K^\prime+P^\prime+V_w^\prime-Q^\prime-2R^\prime,\\
        \epsilon_{16}=2(U^\prime+\mu^\prime+\nu^\prime)+J^\prime+K^\prime+L^\prime+2(P^\prime+V_n^\prime+V_w^\prime)-Q^\prime-2R^\prime,\\
        \epsilon_0=(U^\prime-P^\prime)/2.\\
    \end{multline}

\section*{APPENDIX B: BLOCK-DIAGONAL UNRENORMALIZED HAMILTONIAN}

The matrix elements of the block-diagonal unrenormalized 3-site
Hamiltonian in the $\{|\psi_q\rangle\}$ basis are given in
Eq.(\ref{eq:A2}), where
$\langle\psi_q|-\beta\mathcal{H}_{i,j}-\beta\mathcal{H}_{j,k}|\psi_q\rangle=\varepsilon_q+2G$
for the diagonal elements and
$\langle\psi_q|-\beta\mathcal{H}_{i,j}-\beta\mathcal{H}_{j,k}|\psi_{\bar{q}}\rangle=\varepsilon_{q,\bar{q}}$
for the off-diagonal elements:
    \begin{multline}
    \label{eq:A2}
        \varepsilon_1=0,\quad
        \varepsilon_2=\varepsilon_3=\varepsilon_4=\varepsilon_6/2=\mu,\quad
        \varepsilon_5=\varepsilon_7=2\mu+J,\\
        \varepsilon_8=3\mu+2J,\quad
        \varepsilon_9=\varepsilon_{34}/2=\nu+R,\quad
        \varepsilon_{10}=\nu+2R,\\
        \varepsilon_{12}=\varepsilon_{18}=U/2+\mu+\nu+R/2,\\
        \varepsilon_{13}=\varepsilon_{19}=\mu+\nu+P+R,\quad
        \varepsilon_{15}=\mu+\nu+P,\\
        \varepsilon_{16}=\varepsilon_{49}/2=U+\mu+\nu,\\
        \varepsilon_{21}=U/2+2\mu+\nu/2+J+P+(V_n-R)/2,\\
        \varepsilon_{22}=\varepsilon_{28}=U+2\mu+\nu+J+P+V_n-R,\\
        \varepsilon_{24}=U+2\mu+\nu,\quad
        \varepsilon_{25}=2\mu+\nu+2P,\\
        \varepsilon_{27}=U/2+2\mu+\nu+J+P+(V_n-R)/2,\\
        \varepsilon_{30}=U+3\mu+\nu+2J+P+V_n-R,\\
        \varepsilon_{31}=U+3\mu+\nu+2(J+P+V_n-R),\\
        \varepsilon_{33}=2\nu+K+Q+3R,\\
        \varepsilon_{36}=\varepsilon_{42}=U/2+\mu+2\nu+K+P+(V_w+Q+3R)/2,\\
        \varepsilon_{37}=\varepsilon_{43}=U+\mu+2\nu+R,\\
        \varepsilon_{39}=U+\mu+2\nu+K+P+V_w,\quad
        \varepsilon_{40}=\mu+2(\nu+P),\\
        \varepsilon_{45}=\varepsilon_{51}=3U/2+2(\mu+\nu)+J+K+L/2+2P\qquad\quad\\\qquad\qquad\qquad\qquad\qquad\quad+3(V_n+V_w)/2-(Q+3R)/2,\\
        \varepsilon_{46}=\varepsilon_{52}=U+2(\mu+\nu)+J+2P+V_n-R,\\
        \quad\quad\quad\varepsilon_{48}=U+2(\mu+\nu)+K+2P+V_w,\\
        \varepsilon_{54}=2U+3\mu+2(\nu+J)+K+L+3(P+V_n)+2V_w-Q-3R,\\
        \varepsilon_{55}=2U+3\mu+2(\nu+J+P+V_n-R),\\
        \varepsilon_{57}=3\nu+2(K+Q)+4R,\\
        \varepsilon_{58}=\varepsilon_{60}=U+\mu+3\nu+2K+P+V_w+Q+2R,\\
        \varepsilon_{59}=U+\mu+3\nu+2(K+P+V_w),\\
        \varepsilon_{61}=\varepsilon_{63}=2(U+\mu)+3\nu+J+2K+L+3P+2V_n+3V_w\qquad\quad\\\qquad\qquad\qquad\qquad\qquad\qquad\qquad\qquad\qquad\qquad-Q-2R,\\
        \varepsilon_{62}=2(U+\mu)+3\nu+2(K+P+V_w),\\
        \varepsilon_{64}=3(U+\mu+\nu)+2(J+K+L)+4(P+V_n+V_w-Q-R),\\
        \varepsilon_{2,3}=\varepsilon_{6,7}=\sqrt{2}t,\quad
        \varepsilon_{12,15}=\varepsilon_{24,27}=(2t+R)/\sqrt{2},\\
        \varepsilon_{12,18}=(U-R)/2,\\
        \varepsilon_{13,16}=\varepsilon_{25,28}=\varepsilon_{37,40}=\varepsilon_{49,52}=\sqrt{2}(t+R),\\
        \varepsilon_{15,18}=-\varepsilon_{21,24}=R/\sqrt{2},\qquad
        \varepsilon_{21,27}=(U+V_n-R)/2,\\
        \varepsilon_{36,39}=\varepsilon_{48,51}=(2t+Q+3R)/\sqrt{2},\\
        \varepsilon_{39,42}=-\varepsilon_{45,48}=(Q+R)/\sqrt{2},\\
        \varepsilon_{45,51}=(U+L+V_n+V_w-Q-R)/2,\\
        \varepsilon_{58,59}=\varepsilon_{62,63}=\sqrt{2}(t+Q+2R).\qquad\\
    \end{multline}
The matrix elements for the states connected by the exchange of the
outer conduction electrons are obtained by multiplication with the
eigenvalues $u$ of $T_{ik}$. The matrix elements $\eta_{q,\bar{q}}$
that enter the recursion relations via Eq.(\ref{eq:10}) are obtained
by exponentiating the block-diagonal Hamiltonian given here.

\end{document}